\documentclass[conference]{IEEEtran}

\IEEEoverridecommandlockouts

\usepackage{cite}
\usepackage[T1]{fontenc}
\usepackage[cmintegrals]{newtxmath}
\usepackage{array}
\usepackage[caption=false,font=footnotesize]{subfig}
\usepackage{url}

\usepackage{balance}
\usepackage{graphicx}
\usepackage{algorithm}
\usepackage{algpseudocode}
\usepackage{pifont}
\usepackage{amsmath}
\usepackage{float}
\usepackage{multicol}
\usepackage{mwe}
\usepackage{epstopdf}
\usepackage[utf8]{inputenc}
\DeclareGraphicsExtensions{.eps}
\usepackage{array}
\usepackage{amsmath}
\usepackage{mathtools}

\makeatother
\begin{document}

\title{Backhaul-aware Robust 3D Drone Placement \\in 5G+ Wireless Networks}

\author{\IEEEauthorblockN{Elham Kalantari\IEEEauthorrefmark{1},
		Muhammad Zeeshan Shakir\IEEEauthorrefmark{2},
		Halim Yanikomeroglu\IEEEauthorrefmark{3}, and 
		Abbas Yongacoglu\IEEEauthorrefmark{1}}
	\IEEEauthorblockA{\IEEEauthorrefmark{1}School of Electrical Engineering and Computer Science\\
		University of Ottawa, Ottawa, ON, Canada, Email: \{ekala011, yongac\}@uottawa.ca}
	\IEEEauthorblockA{\IEEEauthorrefmark{2}School of Engineering and Computing\\
	University of the West of Scotland, Paisley, Scotland, UK, Email: muhammad.shakir@uws.ac.uk}
	\IEEEauthorblockA{\IEEEauthorrefmark{3}Department of Systems and Computer Engineering\\
	Carleton University, Ottawa, ON, Canada, Email: halim@sce.carleton.ca}}

\maketitle

\begin{abstract}
Using drones as flying base stations is a promising approach to enhance the network coverage and area capacity by moving supply towards demand when required. However deployment of such base stations can face some restrictions that need to be considered. One of the limitations in drone base stations (drone-BSs) deployment is the availability of reliable wireless backhaul link. This paper investigates how different types of wireless backhaul offering various data rates would affect the number of served users. Two approaches, namely, network-centric and user-centric, are introduced and the optimal 3D backhaul-aware placement of a drone-BS is found for each approach. To this end, the total number of served users and sum-rates are maximized in the network-centric and user-centric frameworks, respectively. Moreover, as it is preferred to decrease drone-BS movements to save more on battery and increase flight time and to reduce the channel variations, the robustness of the network is examined as how sensitive it is with respect to the users displacements.
\end{abstract}


\IEEEpeerreviewmaketitle

\section{Introduction}
\begin{figure*}[t]
	\begin{center}
		\includegraphics[width=6in]{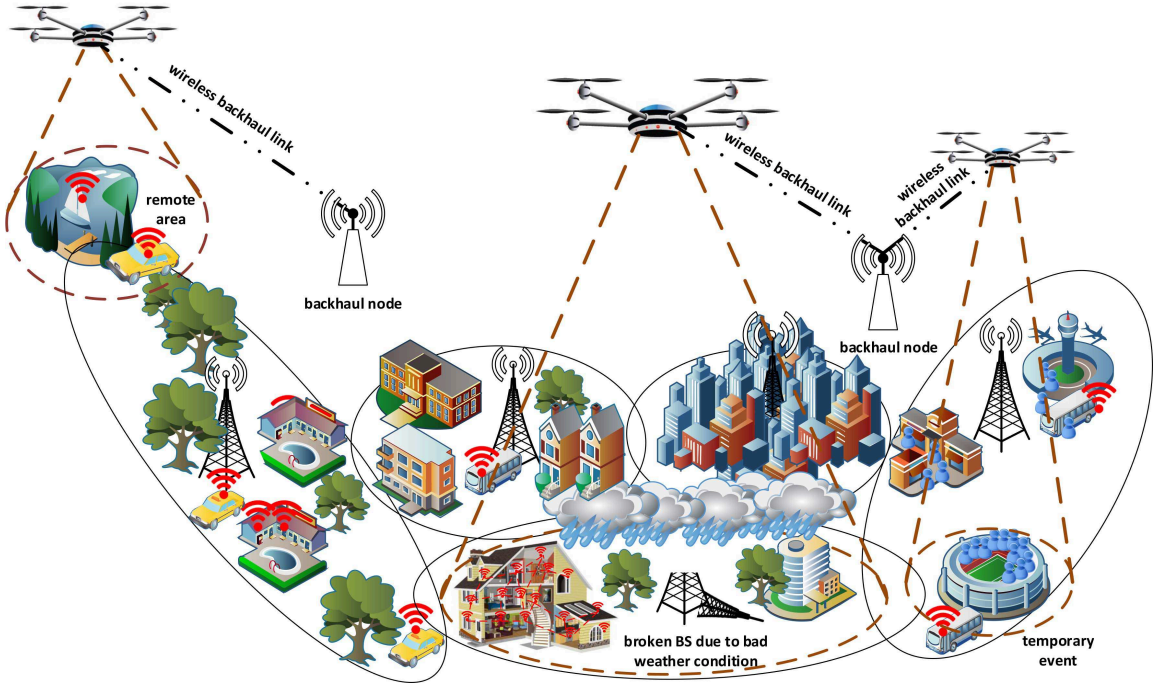}\\
	\end{center}
	\caption{Graphical illustration of various use cases for integration of drone-BSs in cellular networks. Drone-BSs can be exploited in temporary events when the ground-BSs can not serve all the users due to congestion, or when the ground-BS is broken due to bad weather conditions, vandalism, transmission fault, etc. Drone-BSs can also support users in remote areas that there is no coverage by terrestrial networks. The coverage of ground-BSs and drone-BSs are shown by solid-black ellipses and dashed-brown ellipses, respectively.}
	\label{illustrative}
\end{figure*}

Utilization of drone base stations (drone-BSs) in wireless cellular networks has recently attracted a lot of attention as a promising solution to temporarily increase capacity or coverage of an area in 5G+ networks. Drone-BSs can assist a ground network of BSs in providing high data rate coverage whenever and wherever there is an excessive need, especially in situations when this demand occurs in a rather difficult-to-predict manner \cite{eli}. Due to fast deployment of drone-BSs they can also address temporary coverage issues in remote or sparsely populated areas, or when terrestrial wireless infrastructure is damaged due to a natural disaster. Fig. \ref{illustrative} is an illustrative diagram representing some use cases of drone-BSs in future networks. As depicted in this figure, a drone-BS can assist ground network of base stations to inject capacity and prevent temporary congestion in places such as stadiums. It can also provide additional coverage in remote areas or when the ground base stations are out of order due to inclement weather conditions, vandalism, transmission problems, etc.

\subsection{Related Works}
There are a growing number of papers related to integration of drone-BSs in cellular networks discussing drone-BS placement, various use-cases, and design and management challenges. In \cite{iremmag}, a novel framework of multi-tier drone-BSs complementing terrestrial heterogeneous networks (HetNets) is envisioned, and advancements and challenges related to the operation and management of drone-BSs are discussed. In \cite{7470932}, design and implementation challenges of an aerial network of base stations is reported and the capabilities of different aerial platforms for carrying wireless communication systems is reviewed. In \cite{mohammadmag}, a vertical backhaul/fronthaul framework is suggested for transporting the traffic between the access and core networks in a typical HetNet through free space optical (FSO) links. 3D placements of drone-BSs is considered as one of the important problems to design and implement drone-BS enabled HetNets.  There are a few works related to placement of drone-BSs in wireless cellular networks. In \cite{eli}, the authors find the minimum number of drone-BSs and their 3D placement to cover a number of users with high data rate requirement through a heuristic algorithm. They find that in a dense area, a drone-BS will decrease its height to make less interference for farther users that are not served by it, and in a low density region, it will increase its altitude to cover a larger area and serve more users. In \cite{iremconf}, the authors find 3D placement of a drone-BS to maximize the number of covered users through numerical methods. In \cite{6863654}, a closed-form expression for the probability of line-of-sight (LoS) connection between an aerial platform and a receiver is developed and through an analytical approach the optimum altitude that maximizes the radio coverage is obtained. In \cite{MozaffariSBD15a}, the optimal altitude of a drone-BS that achieves a required coverage with minimum transmit power is found. Also providing maximum coverage with two drone-BSs in the presence and absence of interference is investigated. Reference \cite{7486987}, derives the downlink coverage probability of a drone-BS as a function of the altitude and the antenna gain and then determines the locations of drone-BSs in such a way that the total coverage area is maximized. Despite all these recent research, wireless backhaul between the drone-BSs and the core network, is not considered yet as a limiting factor in design and implementation of drone-BS enabled HetNets.

\subsection{Our Contribution}
The major difference between a ground-BS and a drone-BS is that the latter one has a major limitation in the backhaul link. A ground-BS usually has a fixed wired/wireless backhaul connection and can relatively offer very high data rates to a core network. A drone-BS on the other hand, should have a wireless backhaul; therefore, the peak data rate a drone-BS can support is limited and it may dramatically decrease due to inclement weather conditions especially if the link is based on the FSO or mmWave technology. Therefore, an important issue, that to the best of our knowledge has not been addressed yet, is to consider the limitations and requirements of the wireless backhaul link as one of the constraints in designing and deploying the drone-BSs in future 5G+ networks.

The main contribution of this paper is twofold:
\begin{itemize}
	\item We propose a backhaul limited optimal drone-BS placement algorithm for various network design parameters, such as the number of the served users or the sum-rate of the  served users for heterogeneous rate requirements in a clustered user distribution. 
	\item We investigate the robustness of drone-BS placement and study how much the users movements may affect the proposed optimal solution.
\end{itemize}	
The rest of this paper is organized as follows. In Section II, the system model is presented. The optimal drone-BS placement for different design parameters is described in Section III, followed by detailed presentation of the results and related discussions. Finally, conclusions are drawn in Section IV. 

\section{System Model}

\subsection{Pathloss Model}
There are a limited number of studies related to air-to-ground pathloss model. Here, we adopt the one presented in \cite{7037248}. That study shows that there are two main propagation groups, corresponding to the receivers with LoS connections and those with non-line-of-sight connections (NLoS) which can still receive the signal from transmitter due to strong reflections and diffractions. Probability of having a LoS connection between a transmitter and a receiver is an important factor in modeling such channels and it is formulated as \cite{7037248}, \cite{6863654}
\begin{equation}\label{LoS}
P(\textrm{LoS}) = \frac{1}{1+a \exp (-b(\frac{180}{\pi}\theta -a))},
\end{equation}
where $a$ and $b$ are constant values depending on the environment (rural, urban, etc.) and $\theta$ is the elevation angle equal to $\arctan (\frac{h}{r})$, where $h$ and $r$ are the altitude of a drone-BS and its horizontal distance from the receiver, respectively. In this model, shadowing is not considered; instead the average pathloss is presented in a probabilistic manner as \cite{6863654}
\begin{eqnarray}
\nonumber
\textsf{PL}(\textrm{dB}) =& 20 \log(\frac{4\pi f_c d}{c}) ~~~~~~~~~~~~~~~~~~~~~~\\
&+~P(\textrm{LoS})\eta_{LoS}+P(\textrm{NLoS})\eta_{NLoS},
\end{eqnarray}
where the first term is free space pathloss (FSPL) according to Friis equation. Variable $f_c$ is the carrier frequency, $c$ stands for the speed of light, $d$ stands for the distance between a drone-BS and a user and is equal to $\sqrt{h^2+r^2}$. $P(\textrm{NLoS})=1-P(\textrm{LoS})$, and $\eta_{LoS}$ and $\eta_{NLoS}$ are the average additional losses for LoS and NLoS connections, respectively, the values of which depend on the respective environment.

\subsection{Spatial Users Distribution}
To obtain heterogeneity in spatial user distribution, we utilize a $Mat\acute{e}rn$ cluster process \cite{martin-haenggi,5208529}. It is a doubly Poisson cluster process, where \textit{parent} points which are the center of clusters are created by a homogeneous Poisson process. The \textit{daughter} points, that represent users in our model, are uniformly scattered in circles with radius $\nu$ around \textit{parent} points by using another homogeneous spatial Poisson process. Thus, the density function,  $f(z)$, of a given user in location $z$ is
\begin{equation}
f(z)=
\begin{dcases}
\frac{1}{\pi\nu^2}, & \text{if } \|z\|\le \nu,\\
 0,              & \text{otherwise}.
\end{dcases}
\end{equation}

\vspace{4mm}
\section{Backhaul-Aware Drone-BS Placement}
We assume that an area is already covered by ground-BSs, but due to an extensive temporal increase in the number of users or their required rates, some of them can not be served by the terrestrial network due to the lack of resources such as bandwidth. We propose to integrate drone-BS with the existing cellular network infrastructure that offers coverage to such users. The decision about which users to serve in the network, is based on the chosen approach, whether it is network-centric or user-centric. The users are assumed to operate different applications with a variety of rate requirements. The total bandwidth of the drone-BS and the wireless backhaul peak rate are the limiting factors in our formulation.

For the backhaul constraint, we assume that the peak aggregate rate that the wireless backhaul link of a drone-BS can support is $R$ Mbps; so,
\begin{equation}
\sum_{i=1}^{N_U}r_i\cdot I_i\le R,
\end{equation}
where $N_U$ stands for the total number of users that are not served by the terrestrial network, $r_i$ denotes the data rate required by user $i$, and $I_i$ is the user indicator function defined as
\begin{equation}
I_i=
\begin{dcases}
1, \text{ if user } i \text{ is served by the drone-BS,} \\
0, \text{ otherwise.}
\end{dcases}
\end{equation}

Another limiting factor is the total available bandwidth to the drone-BS. It can be formulated as
\begin{equation}
\sum_{i=1}^{N_{U}}b_i\cdot I_i\le B,
\end{equation}
where $B$ stands for the total bandwidth of the drone-BS, and $b_i$ denotes the bandwidth required by user $i$ which is equal to $\frac{r_i}{\zeta_i}$, where $\zeta_i = \log_2(1+\gamma_i)$ represents the spectral efficiency and  $\gamma_i$ stands for the signal-to-noise ratio (SNR) of user $i$. 

Also, we assume that a user is in the coverage of the drone-BS if its quality of service (QoS) requirement is satisfied. It can be formulated as
\begin{equation}
\textsf{PL}_i\cdot I_i \le \textsf{PL}_\textsf{max},\forall i, 
\end{equation}
where $\textsf{PL}_i$ stands for the pathloss when the signal is received by user $i$ and $\textsf{PL}_\textsf{max}$ is the maximum pathloss that a user can tolerate before outage based on its QoS requirement. 

Finally, our optimization problem is formulated as follows:
\begin{equation}
\max_{x,y,h, \{I_i\}} \sum_{i=1}^{N_{U}} \alpha_i\cdot I_i \label{obj1}
\end{equation}
subject to: 
\begin{eqnarray}
\sum_{i=1}^{N_U}r_i\cdot I_i&\le& R  \label{rate}\\
\sum_{i=1}^{N_{U}}b_i\cdot  I_i&\le& B \label{bw}\\
\textsf{PL}_i\cdot I_i &\le& \textsf{PL}_\textsf{max},~~ \forall i \label{PL}\\
x_{min}&\le& x ~~ \le ~~ x_{max}\\
y_{min}&\le& y ~~ \le ~~ y_{max}\\
h_{min}&\le& h ~~ \le ~~ h_{max}\\
I_i &\in& \{0,1\}, ~~\forall i \label{UserServe},
\end{eqnarray}
where $x$, $y$, and $h$ are the 3D coordinates of the drone-BS placement. Variables $x_{min}$, $x_{max}$, $y_{min}$, and $y_{max}$ represent the limits of the area coordinates and $h_{min}$ and $h_{max}$ are the minimum and maximum allowed altitudes of the drone-BS, respectively. The maximum height of a drone-BS depends on its type, size, weight, power of the battery, and other features. It may also be limited by regulatory laws. Several organizations such as US Federal Aviation Authority (FAA), transport Canada, and Canadian Aviation Regulation Advisory Council (CARAC) are working to coordinate such laws \cite{drone}. Variable $\alpha_i$ is a coefficient related to user $i$ and it is determined based on the system, whether it is network-centric or user-centric. It also depends on the metric that is used to identify a user's priority. These concepts will be explained in more detail later in this section.

We propose a centralized solution for finding the best 3D placement of a drone-BS by assuming that the global view of the network is available at a central controller. This can be implemented in the presence of the software-defined networking (SDN) architecture which decouples the control plane from the data plane. Using this approach, we find the best 3D placement of a drone-BS that maximizes the number of users served with higher priority through an exhaustive search. In each candidate coordinates of the drone-BS, the problem can be transformed to a binary integer linear programming (BILP) as given below, which can then be solved through the branch-and-bound method:
\begin{equation}
\max_{\{I_i\}} \sum_{i=1}^{N_{U}} \alpha_i\cdot I_i \label{obj}
\end{equation}
subject to:
 
{\centering
  (\ref{rate}), (\ref{bw}), (\ref{PL}), and (\ref{UserServe}).\par}
  
We consider an urban region with a total area of 16 km$^2$. For the user distribution, we suppose that the \textit{parents}, which represent cluster heads, are created by a Poisson process with an average density of $10^{-7}$ per $m^2$ and \textit{daughters}, which represent users, follow another Poisson distribution with an average density of $90$ users per cluster. The cluster radius is taken as 700 meters. The step size to search 3D location of the drone-BS is 100 meters. The urban environment parameters and the simulation parameters are provided in Table~\ref{table1} and \ref{table2}, respectively. We assume that the rate requirement of user $i$, denoted by $r_i$, is randomly selected from $\mathcal{R}$ ($r_i \in \mathcal{R}$) indicated in Table~\ref{table2}. Matlab software is used to carry out the simulations. 
\begin{table}[t!]
	\renewcommand{\arraystretch}{1.3}
	\centering
	\caption{Urban Environment Parameters}\label{table1}
		\begin{tabular}[t]{|c |c |} 
			\hline
			\scriptsize\textbf{Parameter} & \scriptsize\textbf{Value} \\  
			\hline
			$a$ & 9.61  \\ 
			\hline
			$b$ & 0.16  \\
			\hline
			$\eta_{LoS}$ & 1 dB \\
			\hline
			$\eta_{NLoS}$ & 20 dB \\
			\hline
		\end{tabular}
	\bigskip
	\centering
	\caption{Simulation Parameters}\label{table2}
\begin{tabular}[t]{| c | c | c | c |} 
	\hline
	\scriptsize\textbf{Parameter} & \scriptsize\textbf{Value} & \scriptsize\textbf{Parameter} & \scriptsize\textbf{Value} \\ 
	\hline
	$f_c$ & 2 GHz  & $B$ & 15 MHz \\  
	\hline
	$\textsf{PL}_\textsf{max}$ & $120$ dB & $P_t$ & $5$ Watts  \\
	\hline
	$h_{max}$ & $400$ m & $R$ & $80$ Mbps  \\
	\hline
	$\mathcal{R}$ & \multicolumn{3}{ c| }{$\{0.1, 0.5, 1, 1.5, 2\}$ Mbps}\\
	\hline
\end{tabular}
\end{table}

\subsection{Network-Centric versus User-Centric}
The network may select the users based on the network-centric or the user-centric approach. In the network-centric approach, the network tries to serve as many users as possible, regardless of their rate requirements. As a result, the majority of the served users are the ones who need less data rates. In this approach, $\alpha_i$ in (\ref{obj}), is equal to 1 for all the users. In the user-centric approach, values of $\alpha_i$ vary with the users and they are determined based on the priority of users. A large number of existing and future applications may require differentiation among the users and applications; therefore, offering service only to the users with low rates would not be fair. There are different metrics such as the sum-rate, price differentiation, signal strength, and content demand to identify users priorities. These metrics are explained below:

\subsubsection{Sum-Rate}
One method of selecting users is to maximize the total sum-rate. In this way, by setting $\alpha_i$ equal to $r_i$, the users who require higher data rates are given higher priority to access the network. In this paper, we use this metric in the user-centric approach.

\subsubsection{Price Differentiation}
Users may be categorized based on how much they are willing to pay for their subscribed services, for instance, as platinum, gold, and silver users. The platinum users who pay higher, want to be connected to the network under almost every condition, even if their channel is poor or they need high amount of resources. By assigning a large value to $\alpha_i$ to such users, the service provider makes sure that they are served.

\subsubsection{Signal Strength}
The selection of the users can be based on their received signal strength, so the operator first serves the ones who have favorable channel conditions.

\subsubsection{Content Demand}
In content-aware systems, the users who need to access the network urgently based on their required content, are given higher priority.

\begin{figure}[t]
	\centering
	\subfloat[]{\label{user-dist-2D-net-cen}\includegraphics[width=3.8in]{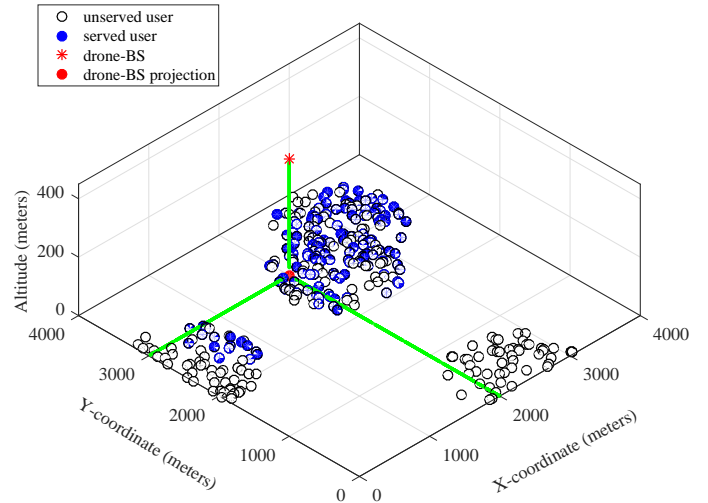}}	
	
	\subfloat[]{\label{user-dist-2D-user-cen}\includegraphics[width=3.8in]{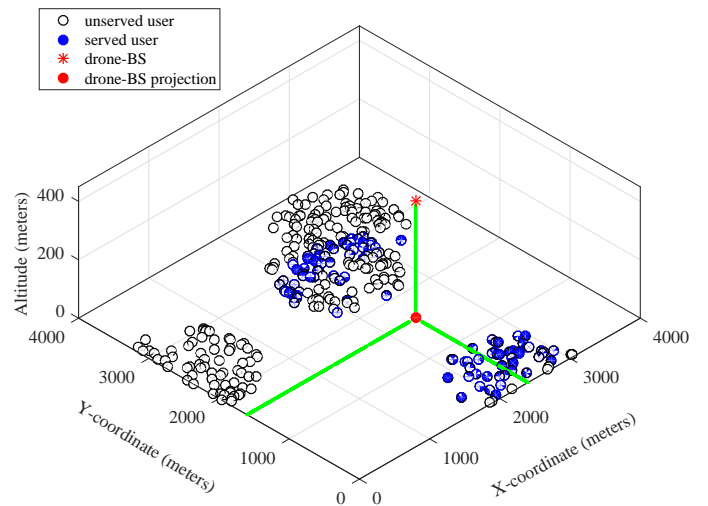}}	
	\caption {User distribution and 3D drone-BS placement in a) network-centric and b) user-centric approaches. The drone-BS and its projection on XY-plane are shown in asterisk and red circles, respectively.}
	\label{user-dist}
\end{figure}

The user distribution and the 3D placement of a drone-BS in a network-centric and a user-centric approach are shown in Fig. \ref{user-dist-2D-net-cen} and \ref{user-dist-2D-user-cen}, respectively. It is observed that in both approaches the drone-BS moves to the highest possible altitude ($h_{max}$) to cover a larger area. As seen in this figure, in the network-centric approach more users are served compared to the user-centric approach. There is a license fee related to spectrum usage that a service provider has to pay which is based on how much bandwidth per person is utilized over a geographical area \cite{website}. Therefore the network-centric approach may be a more favorable option for a service provider as it pays less for the spectrum usage. 

\begin{figure}[t]
	\begin{center}
		\includegraphics[width=3.8in]{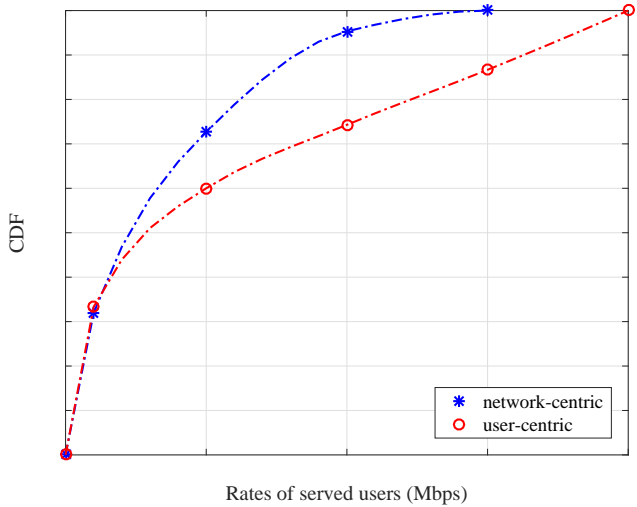}\\
	\end{center}
	\caption{CDF comparison for user-centric and network-centric approaches.}
	\label{CDF-comparison}
\end{figure}

In Fig. \ref{CDF-comparison} the CDF of required rates of the served users for both approaches is depicted. As seen in this figure, the CDF curve related to the network-centric approach is above the user-centric one, meaning that in the former one, there is a higher probability of serving users with lower rates. Therefore, in total more users are served in the network-centric approach as it has been seen earlier in Fig. \ref{user-dist}.

\subsection{Backhaul Limitation}
The backhaul link in a wireless system may be dedicated or in-band.

\subsubsection{Dedicated Backhaul}
Dedicated backhaul may be a FSO or mmWave link between access and core networks. Such links can provide very high backhaul capacity, but they are very sensitive to weather conditions; in foggy or rainy weather, the peak rate may dramatically decrease \cite{mohammadmag}.

\subsubsection{In-band Backhaul}
Currently in LTE, Wi-Fi, WiMAX, and HSPA networks, the main technology used for the wireless backhaul links is based on RF microwave \cite{ericsson}. Microwave backhaul can be deployed very quickly at a relatively low cost. By using RF for backhaul, the same spectrum is used in both the access and backhaul links, so it causes interference and the capacity of the backhaul connection is affected.

\begin{figure}[t]
	\begin{center}
		\includegraphics[width=3.8in]{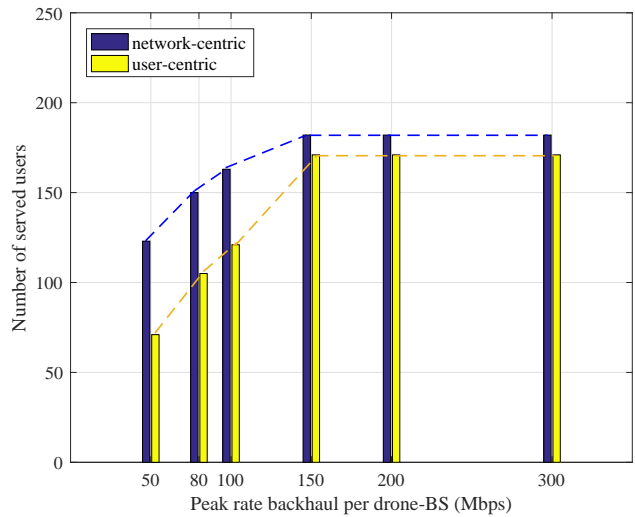}\\
	\end{center}
	\caption{Number of served users versus different wireless backhaul rates.}
	\label{diff-backhaul}
\end{figure}

Fig. \ref{diff-backhaul} compares the number of served users versus different wireless backhaul peak rates of a drone-BS in the network-centric and user-centric approaches. This range of wireless backhaul rates represents the various rates of different types of wireless links. As seen in this figure, low backhaul rates can severely limit the number of served users. By increasing the backhaul capacity, the number of served users is increased differently in two scenarios. It will stop increasing when  the backhaul capacity is around 150 Mbps as there is no more spectrum resource in the drone-BS to serve more users. The speed of increase in the number of served users is almost fixed in the user-centric approach (see fixed slope of the yellow dashed line in Fig. \ref{diff-backhaul}), while it is decreasing in the network-centric approach (see decreasing slope of the blue dashed lines in Fig. \ref{diff-backhaul}). The fixed slope in the user-centric approach is due to the fact that in this scenario, high rate users are served first and when wireless backhaul capacity increases, low rate users receive service, so the amount of increase in the number of served users remain fixed. In the network-centric approach, the slope is not fixed, because low rate users are served first in this scenario; therefore, only a few high rate users get service by increasing the backhaul capacity and the amount of increment is reduced in each step of increasing the backhaul capacity.

\subsection{Robustness}
\begin{figure}[t]
	\centering
	\subfloat[]{\label{robustness_num}\includegraphics[width=3.8in]{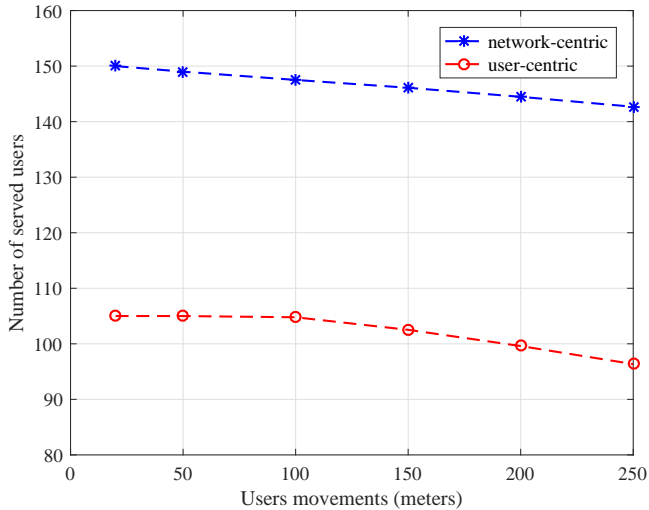}} 
	
	\subfloat[]{\label{robustness_percent}\includegraphics[width=3.8in]{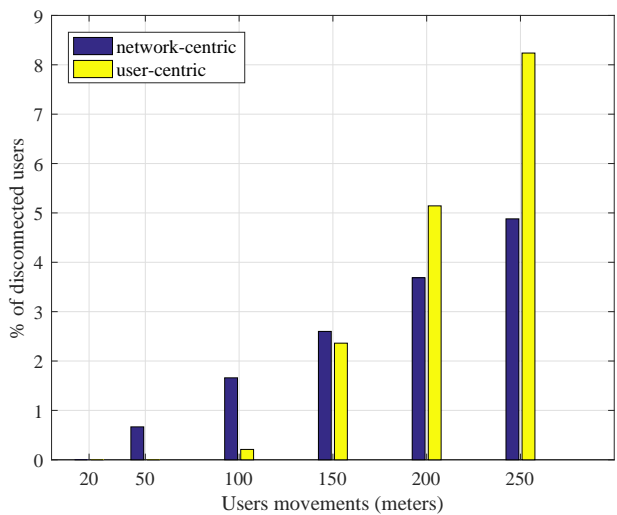}}
	\caption {Impact of user mobility on proposed 3D drone-BS placement in terms of a) number of remaining served users and b) percentage of dropped out users.}
	\label{robustness}
\end{figure}

Mobile drone-BSs change the radio channel persistently, so highly complicated interference management and resource allocation schemes are required. Moreover, constant movements of a drone-BS consume a lot of battery and decrease flight time. Hence, if a drone-BS flies to a predetermined good position and is not required to change its place constantly due to users movements, this will result in savings in energy and reduction in complexity. Fig. \ref{robustness_num} shows the impact of users movements on the performance of the network if the drone-BS stays in its position. As seen, by increasing the movement distance, number of the served users will decrease, but this reduction is not significant and as Fig. \ref{robustness_percent} demonstrates, a very low percentage of users would be dropped out of the network if they move. For instance, if the users are moving within 100 meters, less that 2\% of them in the network-centric approach and less than 1\% in the user-centric approach would be disconnected. Therefore, the solution is robust. If a drone-BS flies to a suitable place, it can stay there for a while unless the network reaches a particular pre-determined user-dropped out threshold.

\section{Conclusion}
In this paper, the optimal 3D placement of a drone-BS over an urban area with users having different rate requirements is investigated. The wireless backhaul peak rate and the bandwidth of a drone-BS are considered as the limiting factors in both the network-centric and user-centric approaches in a typical HetNet. The network-centric approach maximizes the total number of served users regardless of their required rates, while the user-centric approach would maximize their sum-rate. Our investigation also shows that only a small percentage of the total served users would be in outage when the users move. This highlights the robustness of the proposed algorithm against the modest movement of users (within few meters).
\balance

\bibliographystyle{IEEEtran}
\bibliography{Xbib}

\end{document}